\documentclass[a4paper,11pt]{article}
\usepackage{pos}
\usepackage{subcaption}
\usepackage{cleveref} 

\usepackage{multirow}

\def\orcid#1{\kern .08em\href{https://orcid.org/#1}{\includegraphics[keepaspectratio,width=0.7em]{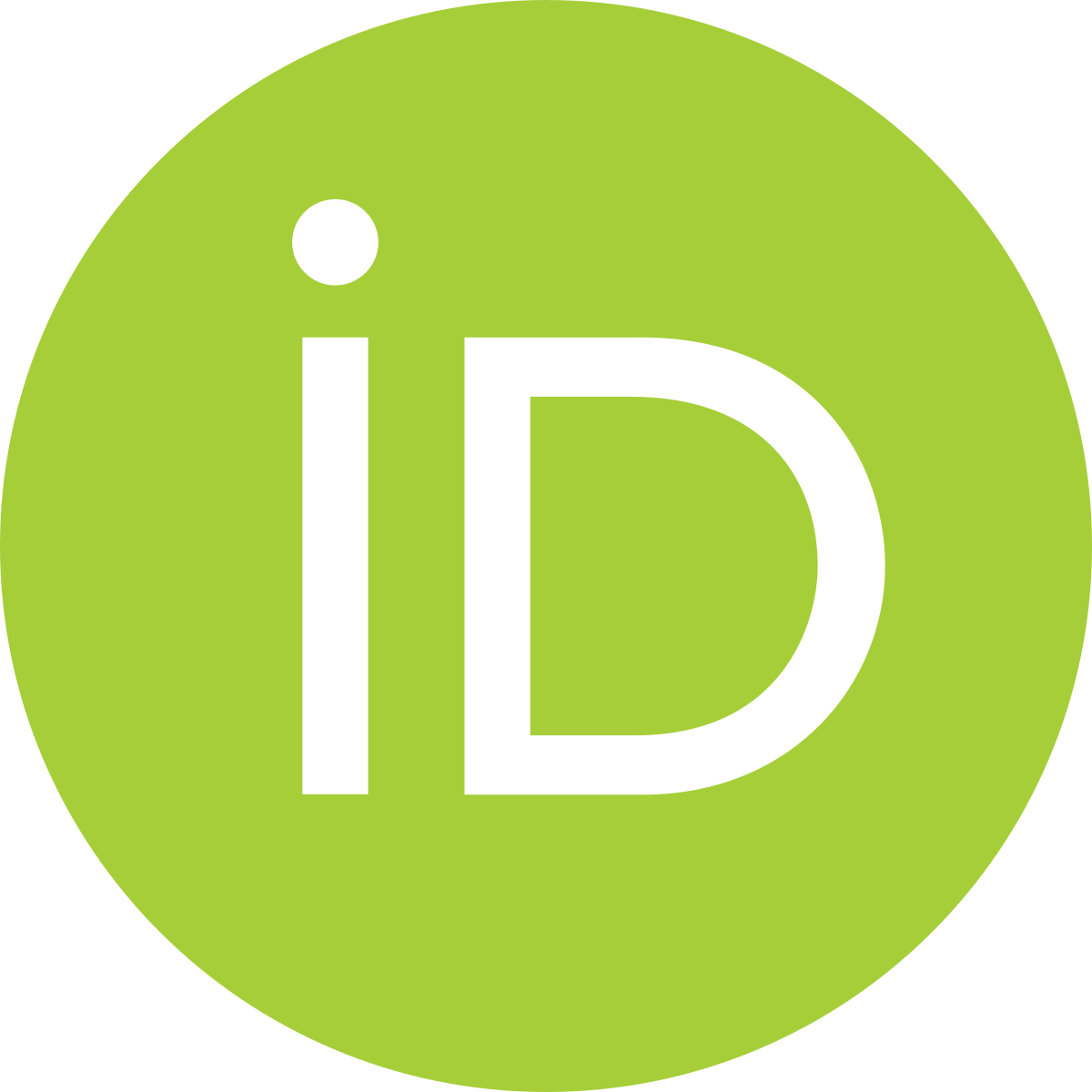}}}

\title{Line shape analysis of $\Lambda(1405)$ in $\gamma p \rightarrow K^+\Sigma^-\pi^+$ reaction using convolutional neural network}
\ShortTitle{Line shape analysis of $\Lambda(1405)$ in $\gamma p \rightarrow K^+\Sigma^-\pi^+$ reaction...}

\author*{Vince Angelo A. Chavez\orcid{0009-0009-2373-1985}}
\author{Denny Lane B. Sombillo\orcid{0000-0001-9357-7236}}

\renewcommand{\printHeadAuthors}{VAA Chavez\orcid{0009-0009-2373-1985} and DLB Sombillo\orcid{0000-0001-9357-7236}}

\affiliation{National Institute of Physics, University of the Philippines Diliman,\\
Quezon City 1101, Philippines}

\emailAdd{vachavez@up.edu.ph}
\emailAdd{dbsombillo@up.edu.ph}

\abstract{Interpreting peaks or dips that appear in an invariant mass distribution is a recurring challenge in hadron physics. These enhancements can be ambiguous, especially near a two-hadron threshold since kinematical and dynamical effects play an important role in their nature. One such enhancement is an exotic baryon $\Lambda(1405)$ which was first observed in 1973. Despite the few available experimental data, the statistics of the measurements of $\Lambda(1405)$ have improved for line shape analysis. The present consensus is that it is a structure of two poles both on the second Riemann sheet. However, there are still investigations of other pole structures corresponding to $\Lambda(1405)$. Lately, the use of a deep neural network in analyzing these line shapes has been proven to be effective, especially in distinguishing pole structures. Thus, in this study, we develop a convolutional neural network, a type of DNN, to determine the general pole structure that corresponds to $\Lambda(1405)$ found in the $\Sigma^-\pi^+$ invariant mass distribution measured by CLAS in their experiment involving the $\gamma p \rightarrow K^+\Sigma\pi$ reaction. The CNN is trained using a two-channel uniformized $S$-matrix allowing us to control the position and the corresponding Riemann sheet of the poles. Our preliminary results show that the trained CNN can accurately distinguish pole structures in the $\Sigma^-\pi^+$ invariant mass distribution and agrees with the present consensus of a two-pole structure. This supports the preceding works on the $\Lambda(1405)$ and requires a thorough analysis of $\Sigma^+\pi^-$ and $\Sigma^0\pi^0$ invariant mass spectra.}

\FullConference{
The 21st International Conference on Hadron Spectroscopy and Structure (HADRON2025)\\
27-31 March, 2025\\
Osaka University, Japan
}

\begin{document}
\maketitle

\section{Introduction}\label{sec:intoduction}
The $\Lambda(1405)$ is widely considered the very first exotic hadron, postulated by Dalitz and Tuan in 1960 \cite{Dalitz:1960du}, then later observed in 1973 \cite{Thomas:1973uh} and 1985 \cite{Hemingway:1984pz}, preceding even the establishment of the quark model. However, its true nature remains a disputed matter to this day. The quantum number assignment of $I(J^P) = 0(1/2^-)$ yields two interpretations for the state of the $\Lambda(1405)$ – that it is an excited quark system of ($uds$) in the $p$-orbital or a possible molecular state of anti-kaon ($\overline{K}$) and nucleon ($N$). Many works are partial to the latter interpretation because of its proximity around the $\overline{K}N$-threshold. Furthermore, several studies have hypothesized that the $\Lambda(1405)$ is a two-pole structure with the first pole being narrow and near the $\overline{K}N$-threshold and the second pole being broad and near the $\Sigma\pi$-threshold -- often attributed to a different state which is the $\Lambda(1380)$ \cite{J-PARCE31:2022plu, Mai:2014xna, Guo:2012vv}. As of now, different techniques have been utilized to verify this two-pole structure.

Machine learning is one of the tools currently used in analyzing these kinds of enhancements. Pioneered in \cite{PhysRevD.102.016024} for single-channel analysis of a nucleon-nucleon scattering, this method has been recently employed in examining the nature of other states such as the $P_{\psi}^{N}(4312)^{+}$ by expanding the approach to two-channel scattering \cite{Santos:2024bqr} and for model-selection framework \cite{Co:2024bfl}. In this study, we applied deep learning to extract the pole structure of  $\Lambda(1405)$. A convolutional neural network (CNN) model was trained and validated using the independent $S$-matrix pole \cite{Santos:2023gfh}, extracting the significant traits of a line shape to be mapped to its corresponding pole structure. We then fed the experimental results from CLAS \cite{CLAS:2014tbc} into the CNN to ascertain the pole structure of $\Lambda(1405)$.

\section{Deep neural network implementation}\label{sec:methodology}

The main objective of this study is to classify line shapes with their corresponding pole structure, such that we may be able to determine the pole structure of the $\Lambda(1405)$ through some training and validation. Some of the key points in the implementation of deep learning in line shape analysis involve (\ref{sec:architecture}) designing the DNN model using building blocks from Pytorch, (\ref{sec:generating dataset}) generating the training and validation datasets via independent $S$-matrix poles through uniformization -- a parametrization of the $S$-matrix, (\ref{sec:training}) training and validating the DNN model to assess and improve its performance in classifying line shapes according to their pole structure, and lastly, (\ref{sec:inference}) using the trained model to give an inference on the pole structure of $\Lambda(1405)$. 


\subsection{Deep neural networks model design} \label{sec:architecture} 

A set of deep neural networks (DNN), discussed in \cite{HORNIK1989359}, is a good universal approximator. One variable can be mapped to another variable through a series of multilayer networks via forward and backward propagation. Previous studies on the implementation of DNN in line shape analysis employed a model with fully connected layers \cite{PhysRevD.102.016024, Santos:2024bqr, Co:2024bfl}. A CNN is a specific type of DNN suited for capturing essential features such as patterns from datasets often in image and video processing \cite{INDOLIA2018679}. In this paper, a CNN is instead used as we wanted to capture features such as the unitarity below the second threshold, and the shape and number of peak structures near a threshold. 

\begin{figure}[tbp]
    \centering
    \subfloat{\includegraphics[height=0.3\linewidth]{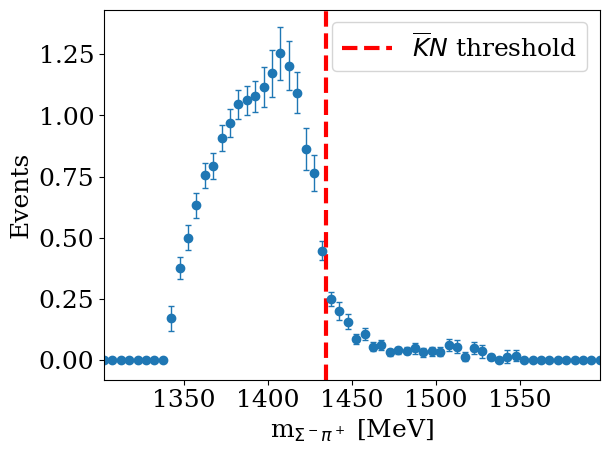}}{\includegraphics[height=0.3\linewidth]{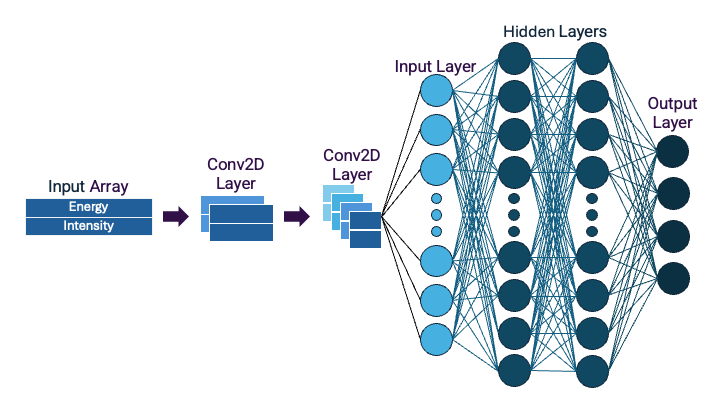}}
    \caption{CLAS measurement \cite{CLAS:2014tbc} of $\Lambda(1405)$ in $\Sigma^-\pi^+$ invariant mass with energy and intensity as inputs for the convolutional neural network model}
    \label{fig:cnn}
\end{figure}

As shown in Fig.\ref{fig:cnn}, our model is comprised of two convolutional layers that extract features from the line shape inputs having output value equal to $z_i = b_i + W_i\star x_i$ where $b$ is the bias, $W$ is the weight, $x$ is the input, and $\star$ is the cross-correlation operator. The features extracted from the convolutional layers will then pass through a series of linear layers that aim to map these features to their corresponding pole structure having a linear function of $z_i=W_{ij}x_j+b_i$. To introduce non-linearity between these linear functions, we included activation layers called Rectified Linear Unit (ReLU) defined as $\text{ReLU(x)}=\max(0,x)$. Lastly, the true and predicted labels of the DNN will then be compared to minimize their difference via a loss function called Cross Entropy Loss.

\subsection{Independent \textit{S}-matrix pole structures as datasets}\label{sec:generating dataset}

After designing the CNN model, we, then, generated the datasets needed for the training of the model. The measurement of CLAS of $\Sigma^-\pi^+$ invariant mass distribution \cite{CLAS:2014tbc} is comprised of energy bins as the independent variable and the number of events or intensities as the dependent variable which suits the characteristics of the line shapes that we are generating. We randomly selected energy values within each energy bin to build the datasets. 

In the paper published by Kato on uniformization \cite{KATO1965130}, a parameterized variable $\omega$ is used to formulate the Jost-like function to construct the $S$-matrix. The variable $\omega$ is derived from the kinematics of the scattering channels through: $\omega=(q_1+q_2)/\sqrt{\epsilon_2^2 - \epsilon_1^2}$ where $q_i$ is the break-up momentum of a two-hadron system in channel $i$, and $\epsilon_i$ is the mass threshold. The independence of the poles is imposed in: $\prod_m D_m(q_1, q_2) = \prod_mD_m(\omega)$, with $D_m(\omega) = (\omega-\omega_m)(\omega-\omega_m^*)(\omega-\omega_{m'})(\omega-\omega_{m'}^*)$ where $\omega_m$ are the relevant poles while $\omega_{m'}$ are the pole regulators. This formulation gives us control over the position of the poles in different Riemann sheets, enabling us to generate multiple datasets with diversified pole structures \cite{Santos:2023gfh}. Thus, this scheme is one of the most appropriate parameterizations for us to use. The $S$-matrix constructed by this formulation is used to extract the $T$-matrix elements for the dataset generating function having the form $F(\sqrt{s}) = |T_{11}(\sqrt{s}) + re^{-i\theta}T_{21}(\sqrt{s})|^2$ where $\sqrt{s}$ is the scattering energy, $T_{11}$ is the $T$-matrix element for scattering channel $\Sigma\pi\rightarrow\Sigma\pi$, $T_{21}$ is the element for scattering channel $\overline{K}N \rightarrow\Sigma\pi$, $r$ is the strength of $T_{21}$, and $e^{-i\theta}$ is the phase factor. 

In Table \ref{tab:pole-config}, we examined 19 pole structures – all of which are configurations with one- to three-pole structures and used as the premise for generating the number of events given the energy values. We generated 10,000 line shapes for each pole structure. The 80\% of these line shapes were used for training and the remaining 20\% for validation. 

\begin{table}[htb]
        \centering
        \caption{Pole structures and corresponding label in the CNN implementation.}
        \label{tab:pole-config}        
        \begin{tabular}{| c | c|| c | c || c | c || c | c |}
            \hline
            \hline
            Label & Pole/s & Label & Pole/s & Label & Pole/s & Label & Pole/s \\
            \hline
            01 & 1 II & 06 & 2 IV & 11 & 3 III & 16 & 2 III, 1 IV \\
            02 & 1 III & 07 & 1 II, 1 III & 12 & 3 IV & 17 & 2 IV, 1 II \\
            03 & 1 IV & 08 & 1 II, 1 IV & 13 & 2 II, 1 III &  18 & 2 IV, 1 III \\
            04 & 2 II & 09 & 1 III, 1 IV & 14 & 2 II, 1 IV & 19 & 1 II, 1 III, 1 IV \\
            05 & 2 III & 10 & 3 II & 15 & 2 III, 1 II & & \\
            \hline
            \hline
        \end{tabular}
    \end{table}


\subsection{Convolutional neural networks model training and validation}\label{sec:training}

In our study, we have modified the curriculum training used in \cite{Santos:2024bqr} to optimize the limited resources that we have. We divided the 19 labels into groups allowing us to expand a single model of 19 output nodes to six (6) different models with only four (4) output nodes shown in Table \ref{tab:training_groupings}. In this way, the accuracy of the model is not compromised by the increasing number of output labels. Through this, we obtained varying models that can classify distinct groups of datasets. 

The first stage is the starting point where we train Labels 1, 2, 3, and 4. The validation was then carried out by measuring different metrics via a confusion matrix such as the precision given by $p = TP/(TP + FP)$ where $TP$ is the true positive referring to the diagonal elements of the confusion matrix, and $FP$ is the false positive referring to the horizontal off-diagonal elements; recall given by $r = TP/(TP + FN)$ where $FN$ is the false negative referring to the vertical off-diagonal elements; and f1-scores given by $f1 = 2pr/(p+r)$ of each label in the model. Afterward, we employed the trained model to obtain the pole structure of $\Lambda(1405)$ using the experimental data measured by CLAS. The next stage would be the training of labels 5, 6, 7, and the inferred label from the previous stage. The process is repeated until it reaches the final stage where, given the total of 19 generated datasets, the inferred label of the last model finally describes the pole structure of $\Lambda(1405)$.

Note that in this analysis, we only considered 19 pole structures. This is not enough to capture all the possible interpretations of these enhancements. However, the modified training technique made the extension of the analysis to structures with more than three poles possible and easier. We can construct another set of DNN models that can capture more complex pole structures without resetting the ones that we already have and compromising the accuracy. 

\begin{table}[htb]
        \centering
        \caption{Groupings of four different labels for the modified curriculum training.}
        \label{tab:training_groupings}        
        \begin{tabular}{| c | c | c | c | c | c | c |}
        \hline
            \hline
            Stages & 1 & 2 & 3 & 4 & 5 & 6 \\
             \hline
             \multirow{4}{*}{Labels} & 01 & Stage 01 Inf & Stage 02 Inf & Stage 03 Inf & Stage 04 Inf & Stage 05 Inf \\
             & 02 & 05 & 08 & 11 & 14 & 17 \\
             & 03 & 06 & 09 & 12 & 15 & 18 \\
             & 04 & 07 & 10 & 13 & 16 & 19 \\
             \hline
             \hline
        \end{tabular}
    \end{table}


\subsection{\texorpdfstring{$\Lambda(1405)$}{Lambda(1405)} pole-structure inference}\label{sec:inference}
After 500 epochs of each stage of training and validation, we proceeded to utilize the model to get an inference on the $\Lambda(1405)$. We input 10,000 line shapes obtained by randomizing the energy and intensity points from the CLAS data into the model. The pole structure with the highest number of predictions is the inference of the model to $\Lambda(1405)$. The inference stage is repeated until all six models are used capturing all of the 19 pole structures generated in Section \ref{sec:generating dataset}.

\section{Results and discussion}\label{sec:results}
\begin{figure}[tbp]
    \centering
        \makebox[1.0\linewidth]{%
        \centering
        \subfloat{\includegraphics[width=0.4\linewidth]{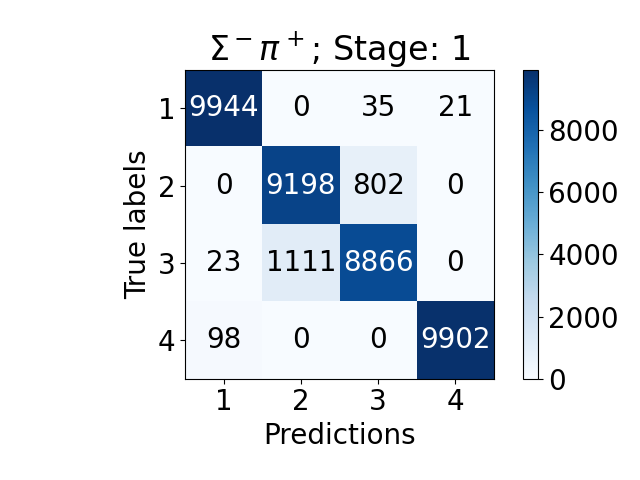}}
        \subfloat{\includegraphics[width=0.4\linewidth]{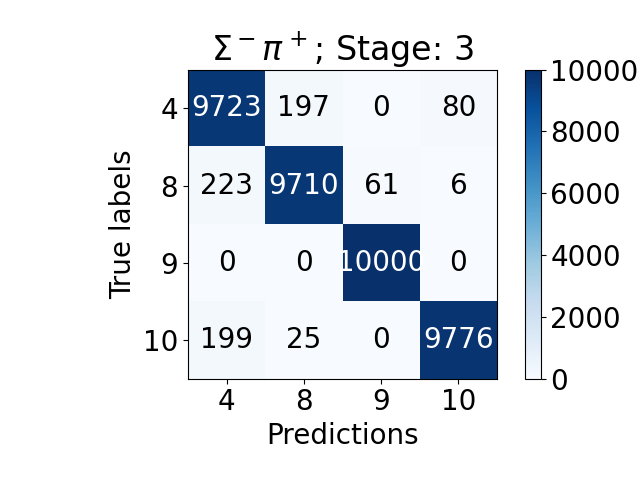}}
        \subfloat{\includegraphics[width=0.4\linewidth]{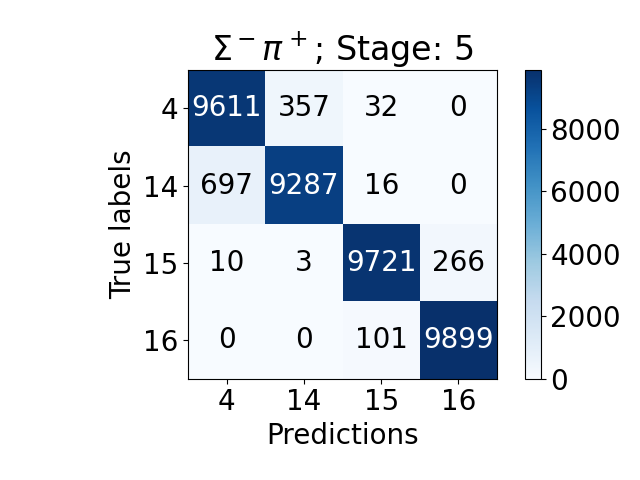}}
  }
    \caption{Confusion matrices from the validation phase of Stages 1, 3 and 5. The labels are in Table \ref{tab:pole-config}.}
    \label{fig:results}
\end{figure}

For Stage 1, we can see that the highest number of predictions per label is located on the diagonal as shown in Fig.\ref{fig:results}. However, the model has a confusion between Labels 2 and 3. Nevertheless, the model favors label 4 with 99\% confidence as shown in Table \ref{tab:metrics}. For Stages 2 and 3,, the highest predictions per label are located on the diagonal. In these stages, there is no confusion among the labels. Both models infer Label 4 with 100\% and 97\% confidence respectively. For Stages 4 and 5, there is a confusion among Labels 4, 13, and 14. Both models infer Label 4 with 95\% confidence which has a confusion with the other labels. However, the confusion is so minimal that it does not significantly affect the findings. Lastly, Stage 6 has the same case as Stage 1 as there is a confusion between Labels 17 and 19. However, this confusion is negligible since the inferred label is Label 4 with 98\% confidence with no confusion with other labels. These results show that among the 19 pole structures used in the analysis, $\Lambda(1405)$ is a two-pole structure in the second Riemann sheet which is consistent with the conclusions of the previous work \cite{J-PARCE31:2022plu, Mai:2014xna, Guo:2012vv}.

\begin{table}[htb]
    \centering
    \caption{Precision, recall, and f1-score of each label in every stage.}
    \label{tab:metrics}
    \makebox[\linewidth]{
    \begin{tabular} {| c | c | c | c | c || c | c | c | c | c |}
        \hline
        \hline
           Stage & Labels & Precision & Recall & F1-Score & Stage & Labels & Precision & Recall & F1-Score \\
        \hline
        \multirow{4}{*}{1} & 01 & 0.99 & 0.99 & 0.99 &\multirow{4}{*}{4} & 04 & 0.93 & 0.96 & 0.95 \\
        & 02 & 0.89 & 0.92 & 0.91 & & 11 & 1.00 & 0.99 & 1.00 \\
        & 03 & 0.91 & 0.89 & 0.90 & & 12 & 0.99 & 1.00 & 0.99 \\
        & 04 & 1.00 & 0.99 & 0.99 & & 13 & 0.96 & 0.92 & 0.94 \\ 
        \hline
        \multirow{4}{*}{2} & 04 & 1.00 & 0.99 & 1.00 & \multirow{4}{*}{5} & 04 & 0.93 & 0.96 & 0.95 \\
        & 05 & 0.97 & 0.96 & 0.96 & & 14 & 0.96 & 0.93 & 0.95 \\
        & 06 & 0.95 & 0.97 & 0.96 & & 15 & 0.98 & 0.97 & 0.98 \\
        & 07 & 0.99 & 0.99 & 0.99 & & 16 & 0.97 & 0.99 & 0.98 \\ 
        \hline
        \multirow{4}{*}{3} & 04 & 0.96 & 0.97 & 0.97 & \multirow{4}{*}{6} & 04 & 0.98 & 0.98 & 0.98 \\
        & 08 & 0.98 & 0.97 & 0.97 & & 17 & 0.81 & 0.83 & 0.82 \\
        & 09 & 0.99 & 1.00 & 1.00 & & 18 & 0.98 & 0.98 & 0.98 \\
        & 10 & 0.99 & 0.98 & 0.98 & & 19 & 0.83 & 0.80 & 0.81 \\ 
        \hline
        \hline
        
    \end{tabular}
    }
\end{table}

\section{Conclusions}\label{sec:conclusions}
In this work, we employed a convolutional neural network in pure line shape analysis of $\Lambda(1405)$ in the $\Sigma^-\pi^+$ invariant mass spectrum. We trained a CNN model using independent $S$-matrix poles, extracting prominent features of a line shape. We then utilized the trained model to get the pole structure of $\Lambda(1405)$. Results show that among the 19 pole structures, the $\Lambda(1405)$ is indeed a two-pole structure in the second Riemann sheet, agreeing with the present consensus.  For future works, we will broaden the analysis by including the $\Sigma^+\pi^-$ and $\Sigma^0\pi^0$ invariant mass spectra, and find the physical interpretation of the pole structure by some model-dependent approach. We will also explore randomized label groupings to address the concern of the possible bias in the current label combinations in the modified training technique.

\section*{Acknowledgments}
VAAC acknowledges the scholarship support provided by the DOST-ASTHRDP.

\bibliographystyle{JHEP}
\bibliography{bib_Chavez}

\end{document}